\documentclass[%
 reprint,
superscriptaddress,
 amsmath,amssymb,
 aps,
 showkeys,
]{revtex4-2}

\usepackage{graphicx}
\usepackage{dcolumn}
\usepackage{bm}
\usepackage{amsthm,mathtools,braket}
\everymath{\displaystyle}

\usepackage[mathlines]{lineno}
\modulolinenumbers[5]

\begin{document}

\preprint{AAPM/123-QED}

\title[Hamiltonian simulation for nonlinear partial differential equation by Schr\"{o}dingerization]{Hamiltonian simulation for nonlinear partial differential equation by Schr\"{o}dingerization}%

\author{Shoya Sasaki}
\affiliation{Department of Science for Open and Environmental Systems, Keio University, 3-14-1 Hiyoshi, Kohoku-ku, Yokohama, Kanagawa 223-8522, Japan}
\author{Katsuhiro Endo}%
\affiliation{%
 National Institute of Advanced Industrial Science and Technology (AIST),
 Research Center for Computational Design of Advanced Functional Materials,
 Central 2, 1-1-1 Umezono, Tsukuba, Ibaraki 305-8568, Japan
}%
\author{Mayu Muramatsu}
\email{muramatsu@mech.keio.ac.jp}
\affiliation{%
Department of Mechanical Engineering, Keio University, 3-14-1 Hiyoshi, Kohoku-ku, Yokohama, Kanagawa 223-8522, Japan
}%


\begin{abstract}
Hamiltonian simulation is a fundamental algorithm in quantum computing that has attracted considerable interest owing to its potential to efficiently solve the governing equations of large-scale classical systems.
Exponential speedup through Hamiltonian simulation has been rigorously demonstrated in the case of coupled harmonic oscillators.
The question arises as to whether Hamiltonian simulations in other physical systems also accelerate exponentially.
Schr\"{o}dingerization is a technique that transforms the governing equations of classical systems into the Schr\"{o}dinger equation.
However, since the Schr\"{o}dinger equation is a linear equation, Hamiltonian simulation is often limited to linear equations.
The research on Hamiltonian simulation methods for nonlinear governing equations remains relatively limited.
In this study, we propose a Hamiltonian simulation method for nonlinear partial differential equations (PDEs).
The proposed method is named Carleman linearization + Schr\"{o}dingerization (CLS), which combines Carleman linearization (CL) and warped phase transformation (WPT).
CL is first applied to transform a nonlinear PDE into a linear differential equation. 
This linearized equation is then mapped to the Schrödinger equation via WPT.
The original nonlinear PDE can be solved efficiently by the Hamiltonian simulation of the resulting Schrödinger equation.
By applying this method, we transform the original governing equation into the Schrödinger equation. Solving the transformed Schrödinger equation then enables the analysis of the original nonlinear equation.
As a specific application, we apply this method to the nonlinear reaction--diffusion equation to demonstrate that Hamiltonian simulations are applicable to nonlinear PDEs.
\end{abstract}

\maketitle

\section{Introduction} \label{sec:introduction}
Partial differential equations (PDEs) describe a wide range of physical phenomena, such as heat conduction, microstructure evolution in materials, and fluid dynamics.
PDE-based analysis plays an important role in analyzing physical phenomena in the real world.
One major challenge in PDE-based analysis is the difficulty in solving PDEs for extremely large-scale systems within practical time frames \cite{Sato2024scalableqc,Sato2024LCHS}.
Quantum computing is a promising approach to overcoming this challenge. 
It has been increasingly attracting attention owing to its potential to accelerate the large-scale analysis of PDEs \cite{Gibney2019quantumsupremacy}.
Quantum computing utilizes fundamental principles of quantum mechanics, such as superposition and entanglement, to perform calculations \cite{Sukulthanasorn2025}.
Compared with classical computing, quantum computing provides significant advantages in the analysis of large-scale PDEs \cite{LiuOrtiz2023,Sarma2024}.
Various methodologies have been investigated for simulating the time evolution of PDEs by quantum computing.
These approaches can be broadly classified into two main categories.

The first category includes methods of solving difference equations obtained by discretizing time-dependent PDEs in the time and space directions using matrix operations.
The first category involves the application of quantum algorithms originally developed for linear algebra problems \cite{Costa2022,Motlagh2023}, such as the quantum linear systems algorithm (QLSA) \cite{Childs2020,Berry2017,Berry2014,Childs2017} including the Harrow--Hassidim--Lloyd algorithm \cite{Harrow2009HHL,Dervovic2018,Childs2017HHLOPTIMAL,Ye2023,Pilaszewicz2025}.

The second category includes Hamiltonian simulation methods \cite{Low2017,Berry2015,Childs2018}.
The task of solving the time evolution of the solution to a Schrödinger equation for a time-independent Hamiltonian is called the Hamiltonian simulation problem \cite{Linlinlecture2022}.
The Hamiltonian simulation problem is rewritten in short as follows: give an initial state and Hamiltonian, and then find the time evolution of the solution.
Since the Hamiltonian determines the time evolution of a system, the time evolution of various physical systems can be obtained by Hamiltonian simulation through the design of the Hamiltonian.
Babbush et al. \cite{Babbush2023quantumspeedup} rigorously demonstrated an exponential quantum speedup by Hamiltonian simulation to a system of classical harmonic oscillators.
Hamiltonian simulation is attracting attention as a method that has the potential to accelerate PDE-based analysis.

An analysis method based on nonlinear PDEs is also important because many real-world phenomena are nonlinear such as large deformation in materials, turbulence in fluid flow, chaotic systems and reaction--diffusion phenomena.
The Hamiltonian simulation of nonlinear PDEs is difficult, and there has not been sufficient discussion or verification yet.
There are two key challenges in performing Hamiltonian simulations of nonlinear PDEs.

Firstly, the target equation is a nonlinear PDE, whereas a Schrödinger equation is a linear PDE.
This means that the target nonlinear PDEs must be transformed into a linear equation.
Some algorithms for linearization that can be realized in quantum computing have been proposed.
Joseph \cite{Joseph2020} considered Koopman von Neumann (KvN) linearization based on the Koopman operator \cite{Koopman1931} in quantum computing.
KvN linearization is a general linearization method with a high degree of freedom in basis functions.
Liu and coworkers \cite{Liu2021,Liu2023carlemanreactiondiffusion} applied Carleman linearization (CL) \cite{Liu2021,Liu2023carlemanreactiondiffusion,Amini2025,Akiba2023,Forets2021,Forets2017,Sanavio2024,Shi2024} to linearize the nonlinear reaction--diffusion equation and employed QLSA to compute physical quantities such as energy.
CL is a linearization method used when selecting polynomials as basis functions in KvN.
Endo and Takahashi \cite{Endo2024} proposed an algorithm that mitigates the divergence of solutions caused by CL when implemented on a quantum computer.
In this study, we focused on CL as a linearization method because it is a fundamental linearization method and has been extensively studied for its applications in quantum computing.

\begin{figure*}[t]
  \centering
  \includegraphics[width=\linewidth]{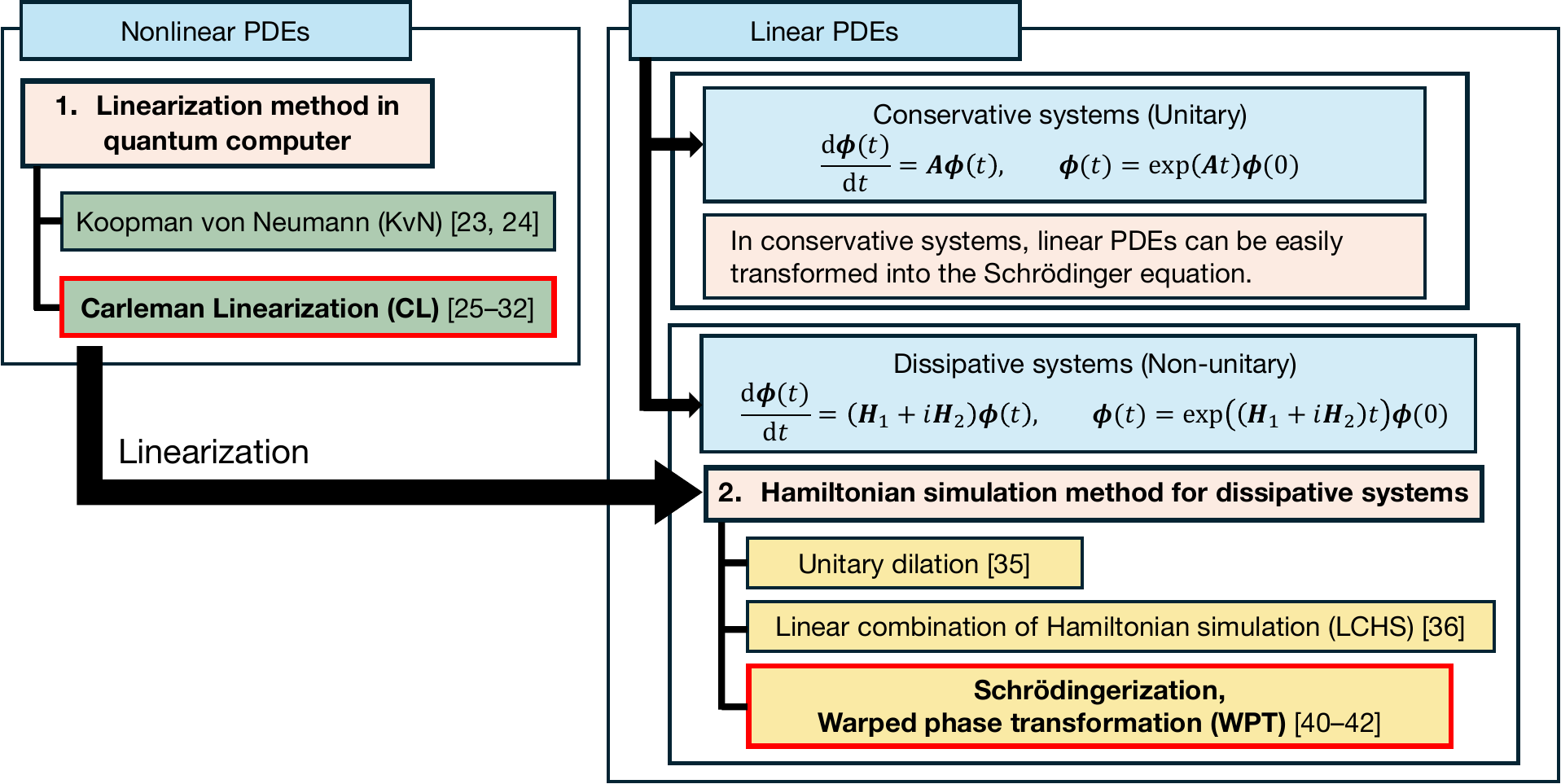}
  \caption{Two key challenges to performing Hamiltonian simulations of nonlinear PDEs and classification of previous methods for Hamiltonian simulation.}
  \label{fig:classification_diagram}
\end{figure*}

Secondly, a linearized equation is generally a dissipative system, whereas a Schrödinger equation is a conservative system.
In this study, a conservative system is defined as the system with the time evolution operator represented by a unitarity operator. 
In contrast, a dissipative system is defined as the system with the time evolution operator represented by a non-unitarity operator.
In conservative systems, linear PDEs can be easily transformed into a Schrödinger equation.
Costa et al. \cite{Costa2019} proposed a quantum algorithm for simulating the wave equation under Dirichlet and Neumann boundary conditions, using Hamiltonian simulation as a subroutine.
Sato et al. \cite{Sato2024scalableqc} proposed a method of explicitly implementing quantum circuits for Hamiltonian simulation, applied it to linear advection and wave equations, and highlighted its potential for exponential speedup.
However, these studies have been limited to conservative systems.
Several methods have been proposed for handling dissipative systems on quantum computers.
Gonzalez-Conde et al. \cite{Gonzalez-Conde2023} converted the Black--Scholes equation in a dissipative system into the Schr\"{o}dinger equation in a conservative system by unitary dilation.
Unitary dilation introduces an additional ancilla qubit to the system, allowing the time evolution operator to be unitary.
An et al. \cite{An2023} proposed the linear combination of Hamiltonian simulation (LCHS) as an approach to handling dissipative systems.
LCHS can be viewed as a special case of linear combination of unitaries (LCU) \cite{Childs2012,Childs2021,Meister2022}.
Jin and coworkers \cite{Jin2022,Jin2023schodingerizaiton-first,Jin2024schrodngerization-conditions} proposed Schr\"{o}dingerization, which is a method for mapping a general linear ordinary differential equation (ODE) including a dissipative system to a Schrödinger equation.
The core of Schrödingerization is warped phase transformation (WPT), which converts a dissipative system into a conservative system by adding new auxiliary variables to the spatial dimensions of the system.
In this study, we focused on WPT because unitary dilation and LCHS can only succeed in Hamiltonian simulation probabilistically, but WPT does not require probabilistic Hamiltonian simulation.
However, it is unclear what advantages there are when postselection is included.
In Fig.~\ref{fig:classification_diagram} two key challenges to performing Hamiltonian simulations of nonlinear PDEs are summarized and previous studies are classified.

In this study, we propose a method for the Hamiltonian simulation of nonlinear PDEs, named Carleman linearization + Schr\"{o}dingerization (CLS).
The proposed CLS framework integrates the following two components: first, CL is applied to transform the target nonlinear PDE into a system of linear ODEs.
Then, the resulting linear system is converted into a Schr\"{o}dinger equation using WPT.
Finally, Hamiltonian simulation is applied to the transformed Schr\"{o}dinger equation, allowing the time evolution of the original nonlinear PDE to be recovered.

In this study, we apply CLS to a nonlinear reaction--diffusion equation as a representative example of a nonlinear PDE.  
We evaluate the time evolution of the solution obtained by CLS, assess the associated errors, and verify the computational accuracy, thereby demonstrating the effectiveness of the CLS method.

\section{Theory}
\subsection{Nonlinear reaction--diffusion equation}
The nonlinear reaction--diffusion equation describes the time evolution of a system in which two processes, reaction and diffusion, proceed simultaneously.
These processes include ecology, combustion, phase separation, and tissue formation phenomena.
Letting $t$ denote the time, $\bm{x}$ represent the spatial coordinates, and $\phi(t,\bm{x})$ the field variable, the reaction--diffusion equation is generally given by \cite{DeMasi1986}
\begin{equation}
	\frac{\partial{\phi}(t,\bm{x})}{\partial t}=D\nabla^2{\phi}(t,\bm{x})+f({\phi}(t,\bm{x})),
	\label{eq:reaction_diffusion}
\end{equation}
where $D\in\mathbb{R}_+$ is the diffusion coefficient and $f(\phi(t,\bm{x}))$ is at least the $C^1$ class smooth function.
In Eq.~\eqref{eq:reaction_diffusion}, the term $D\nabla^2\phi$ is the linear component, and $f(\phi(t,\bm{x}))$ corresponds to the nonlinear component.
Therefore, the reaction--diffusion equation can be classified as a type of nonlinear PDE.
In this study, we assume that $f(\phi(t,\bm{x}))$ is expressed as a quadratic function for $\phi$, given by
\begin{equation}
	f(\phi(t,\bm{x}))=Q\phi(t,\bm{x})+R\phi(t,\bm{x})^2,
	\label{eq:reaction}
\end{equation}
where $Q\in\mathbb{R}$ is the coefficient of the first term and $R\in\mathbb{R}$ is the coefficient of the second term.
Substituting Eq.~\eqref{eq:reaction} into Eq.~\eqref{eq:reaction_diffusion}, we obtain the following equation:
\begin{equation}
	\frac{\partial\phi(t,\bm{x})}{\partial t}=D\nabla^2\phi(t,\bm{x})+Q\phi(t,\bm{x})+R\phi(t,\bm{x})^2.
	\label{eq:reaction_diffusion_transformed}
\end{equation}

\subsection{CL}
CL is a linearization technique that transforms a finite-dimensional nonlinear system into an infinite-dimensional linear system by extending the state variables into an infinite-dimensional space.
In this section, we consider the application of CL to a general nonlinear differential equation.
Nonlinear differential equations are generally expressed as
\begin{equation}
    \frac{\mathrm{d}\bm{x}}{\mathrm{d}t}=\bm{f}(t,\bm{x}).
    \label{eq:nonlinear_ode}
\end{equation}
Here, $\bm{x}=[x_1,\ldots,x_n]^{\mathrm{T}}\in\mathbb{R}^n$ is the state vector of the system and $\bm{f}(t,\bm{x}) \in \mathbb{R}^n$ is an analytic function of $\bm{x}$, defined as $\bm{f}:\mathbb{R}\times\mathbb{R}^n\rightarrow\mathbb{R}^n$.
Approximating $\bm{f}(t,\bm{x})$ with a polynomial is given by
\begin{equation}
    \bm{f}(t,\bm{x}(t))=\sum_{m=0}^\infty\bm{F}_m\bm{x}^{\otimes m}=\bm{F}_0+\bm{F}_1\bm{x}+\bm{F}_2\bm{x}^{\otimes 2}+\cdots,
    \label{eq:taylorseries_nonlinearode}
\end{equation}
where $\otimes$ represents Kronecker's product, $\bm{x}^{\otimes m}=\overbrace{\bm{x}\otimes\cdots\otimes\bm{x}}^{m~\text{times}}\in\mathbb{R}^{n^m}$ for any given non-negative integer $m$, and $\bm{F}_m\in\mathbb{R}^{n\times n^m}$ is a coefficient matrix of $\bm{x}^{\otimes m}$.
For notational convenience, $\bm{x}^{\otimes 0}\coloneqq 1$.
The Kronecker product is an operation on two matrices of arbitrary size.
The result of the operation is given as a matrix expanding the set of bases (i.e., $\otimes$: $(\mathbb{R}^{a}\times\mathbb{R}^{b}),(\mathbb{R}^{c}\times\mathbb{R}^{d})\rightarrow(\mathbb{R}^{ac}\times\mathbb{R}^{bd})$).
In CL, we consider the time evolution of extended variables $\bm{y}_k\coloneqq \bm{x}^{\otimes k}$ for any non-negative integer $k$.
According to Eqs. \eqref{eq:nonlinear_ode} and \eqref{eq:taylorseries_nonlinearode}, and applying the chain rule, we obtain the time derivative of $\bm{y}_k$ as
\begin{equation}
    \frac{\mathrm{d}\bm{y}_k}{\mathrm{d}t}=\frac{\mathrm{d}\bm{x}^{\otimes k}}{\mathrm{d}t}=\frac{\mathrm{d}\bm{x}^{\otimes k}}{\mathrm{d}\bm{x}}\frac{\mathrm{d}\bm{x}}{\mathrm{d}t}=\frac{\mathrm{d}\bm{x}^{\otimes k}}{\mathrm{d}\bm{x}}\sum_{m=0}^\infty\bm{F}_m\bm{x}^{\otimes m}.
    \label{eq:carleman_y_k}
\end{equation}
Applying the product rule of differentiation, we calculate $\mathrm{d}\bm{x}^{\otimes k}/\mathrm{d}\bm{x}$ as:
\begin{equation}
    \frac{\mathrm{d}\bm{x}^{\otimes k}}{\mathrm{d}\bm{x}}=\sum_{v=0}^{k-1}\bm{x}^{\otimes v}\otimes \bm{I} \otimes \bm{x}^{\otimes k-1-v}.
    \label{eq:tensor_derivative}
\end{equation}
Substituting Eq.~\eqref{eq:tensor_derivative} into Eq.~\eqref{eq:carleman_y_k}, we obtain
\begin{equation}
    \frac{\mathrm{d}\bm{y}_k}{\mathrm{d}t}=\sum_{m=0}^\infty\left(\sum_{v=0}^{k-1}\bm{I}^{\otimes v}\otimes \bm{F}_m \otimes \bm{I}^{\otimes k-1-v}\right)\bm{x}^{\otimes m+k-1}.
    \label{eq:carleman_infinite_primitive}
\end{equation}
Here, we use the relationship $(\bm{A}\otimes \bm{B})(\bm{C}\otimes \bm{D})=\bm{AC}\otimes\bm{BD}$, which holds true when the matrices $\bm{A}, \bm{B}, \bm{C}$, and $\bm{D}$ are of a size such that the matrix products $\bm{AC}$ and $\bm{BD}$ can be defined.
Here, by letting $l$ be $m + k - 1$, we obtain the time evolution of $\bm{y}_k$ by calculating the following infinite-dimensional linear differential equation:
\begin{equation}
    \frac{\mathrm{d}\bm{y}_k}{\mathrm{d}t}=\sum_{l=0}^\infty\bm{A}_{k,l}\bm{y}_l,
    \label{eq:nonlinear_carleman_infinite}
\end{equation}
where
\begin{equation}
    \bm{A}_{k,l}=\sum_{v=0}^{k-1}\bm{I}^{\otimes v}\otimes \bm{F}_{l-k+1} \otimes \bm{I}^{\otimes k-1-v}.
\end{equation}
Eq.~\eqref{eq:nonlinear_carleman_infinite} is an infinite-dimensional linear differential equation.
Since it is not possible to solve it in an infinite-dimensional space, we truncate Eq.~\eqref{eq:nonlinear_carleman_infinite} at the order of $K$ and compute it as the following approximate finite-dimensional linear ODE:
\begin{equation}
    \frac{\mathrm{d}\bm{y}_k}{\mathrm{d}t}=\sum_{l=0}^{K}\bm{A}_{k,l}\bm{y}_{l},~~~~0\leq k\leq K.
    \label{eq:nonlinear_carleman_finite}
\end{equation}
Note that Eq.~\eqref{eq:nonlinear_carleman_finite} is a dissipative system.
To perform the Hamiltonian simulation of Eq.~\eqref{eq:nonlinear_carleman_finite}, a Hamiltonian simulation framework for dissipative systems is required.

\subsection{Schr\"{o}dingerization using WPT} \label{sec:WPT}
In this section, we introduce Schrödingerization framework which is a method for mapping a general linear ODE including a dissipative system to a Schrödinger equation.
The core of Schrödingerization is WPT.
In WPT, a dissipative system can be transformed into a conservative one by introducing an auxiliary variable independent of the spatial dimensions.
We consider the application of WPT to a general linear ODE.
A linear ODE is generally expressed as
\begin{equation}
    \frac{\mathrm{d}\bm{u}(t)}{\mathrm{d}t}=\bm{A}\bm{u}(t),
    \label{eq:linear_ode}
\end{equation}
where $\bm{u}(t)\in\mathbb{C}^n$ is the state vector in the system and $\bm{A}\in\mathbb{C}^{n\times n}$ is a coefficient matrix.
The WPT of dissipative systems into conservative systems is achieved by introducing an auxiliary variable independent of the spatial dimensions.
First, since the coefficient matrix $\bm{A}$ is a square matrix, it is decomposed into its Hermitian part $\bm{H}_1$ and skew-Hermitian part $i\bm{H}_2$ as
\begin{equation}
    \bm{A}=\bm{H}_1+i\bm{H}_2.
    \label{eq:decomposition_A}
\end{equation} 
The Hermitian part $\bm{H}_1$ and the skew-Hermitian part $i\bm{H}_2$ are defined as
\begin{equation}
  \bm{H}_1=\frac{\bm{A}+\bm{A}^\dagger}{2},\quad i\bm{H}_2=\frac{\bm{A}-\bm{A}^\dagger}{2}.
\end{equation}
In WPT, we introduce an auxiliary variable, $p\geq0$.
WPT is formulated as
\begin{equation}
    \bm{v}(t,p)=e^{-p}\bm{u}(t).
	\label{eq:wpt_transformation}
\end{equation}
Multiplying both sides of Eq.~\eqref{eq:linear_ode} by $e^{-p}$ yields
\begin{equation}
    \frac{\mathrm{d}}{\mathrm{d}t}\left(e^{-p}\bm{u}(t)\right)=e^{-p}\bm{A}\bm{u}(t).
    \label{eq:linear_ode_multiplied_ep}
\end{equation}
Substituting Eq.~\eqref{eq:decomposition_A} into Eq.~\eqref{eq:linear_ode_multiplied_ep} yields
\begin{equation}
  \frac{\mathrm{d}}{\mathrm{d}t}\left(e^{-p}\bm{u}(t)\right)=e^{-p}\left(\bm{H}_1+i\bm{H}_2\right)\bm{u}(t).
  \label{eq:linear_ode_multiplied_ep_subAdecomposition}
\end{equation}
Eq.~\eqref{eq:linear_ode_multiplied_ep_subAdecomposition} can be transformed as
\begin{equation}
  \frac{\mathrm{d}}{\mathrm{d} t}\left(e^{-p}\bm{u}(t)\right)=\left(-\bm{H}_1\frac{\partial}{\partial p}+i\bm{H}_2\right)e^{-p}\bm{u}(t).
  \label{eq:linear_ode_multiplied_ep_subAdecomposition_transformed}
\end{equation}
Substituting Eq.~\eqref{eq:wpt_transformation} into Eq.~\eqref{eq:linear_ode_multiplied_ep_subAdecomposition_transformed} yields
\begin{equation}
  \frac{\mathrm{d}\bm{v}(t,p)}{\mathrm{d}t}=-\bm{H}_1\frac{\partial \bm{v}(t,p)}{\partial p}+i\bm{H}_2\bm{v}(t,p).
  \label{eq:linear_transformed_wpt}
\end{equation}
The first term on the right-hand side of Eq.~\eqref{eq:linear_transformed_wpt} captures the advection of $\bm{v}(t,p)$.
Therefore, Eq.~\eqref{eq:linear_transformed_wpt} should be discretized in the $p$-direction by the upwind difference method.
The upwind difference method is a discretization technique in which the spatial differential term is approximated using the difference between a reference point and an upstream point.
Furthermore, even if the initial value is extended to the region of $p<0$, the solution $\bm{v}(t,p)$ does not impact  the region $p\geq0$, because it flows from right to left in the $p$-direction.
Therefore, we extend Eq.~\eqref{eq:linear_transformed_wpt} to $p<0$ with the following initial condition:
\begin{equation}
  \bm{v}(0,p)=e^{-|p|}\bm{u}(0).
\end{equation}
As a result, the ODE represented by Eq.~\eqref{eq:linear_ode} is transformed into the following system:
\begin{equation}
  \begin{cases}
    \frac{\mathrm{d}\bm{v}(t,p)}{\mathrm{d}t}=\bm{A}'\bm{v}(t,p),\\
    \bm{v}(0,p)=e^{-|p|}\bm{u}(0),
  \end{cases}
  \label{eq:wpt_schrodinger}
\end{equation}
where
\begin{equation}
  \bm{A}'\coloneqq -\bm{H}_1\frac{\partial}{\partial p}+i\bm{H}_2.
  \label{eq:A_prime}
\end{equation}
Note that the matrix $\bm{A}'$ is a skew-Hermitian matrix and the proof is shown in \ref{sec:hermitian}.
Therefore, the time evolution of Eq.~\eqref{eq:wpt_schrodinger} is unitary and Eq.~\eqref{eq:wpt_schrodinger} is a Schr\"{o}dinger equation.
The unitarity of the time evolution of Eq.~\eqref{eq:wpt_schrodinger} implies that Eq.~\eqref{eq:linear_ode} is suitable for Hamiltonian simulation via WPT.

\section{Method}
\subsection{CLS}\label{sec:cls}
\begin{figure*}[t]
    \centering
    \includegraphics[width=\linewidth]{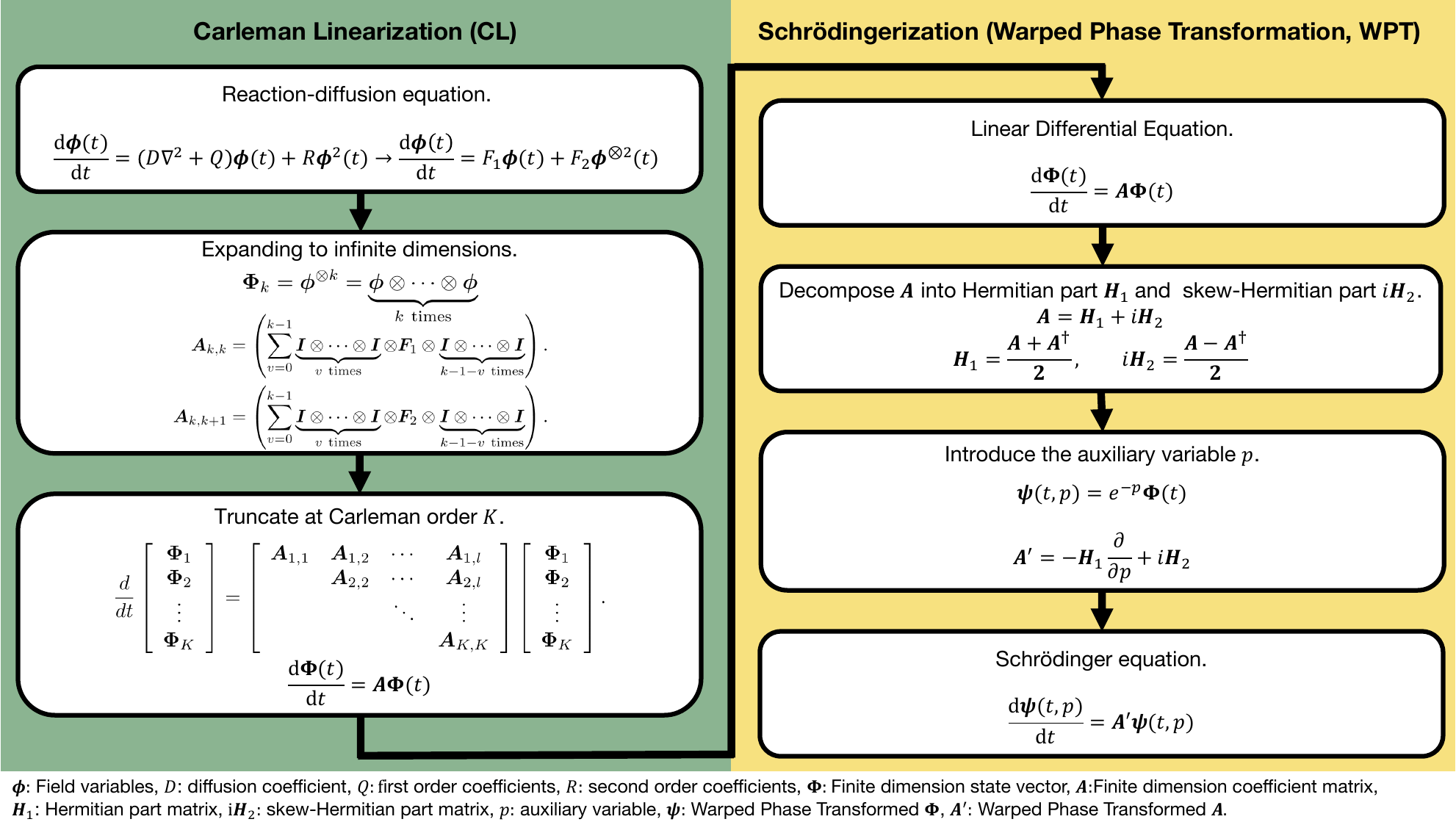}
    \caption{Flow of CLS.}
    \label{fig:carleman_schrodingerization}
  \end{figure*}
In this study, we propose CLS as a Hamiltonian simulation method for nonlinear PDEs.
Specifically, CL is used to transform a nonlinear PDE into a linear ODE, and WPT is used to transform a linearized ODE in a dissipative system into a Schr\"{o}dinger equation.
In this study, we apply CLS to nonlinear reaction--diffusion equations and examine its usefulness.
First, we consider the discretization of the nonlinear reaction--diffusion equation shown in Eq.~\eqref{eq:reaction_diffusion_transformed} in the $x$-direction.
Let $\Omega_x\coloneqq(0, x_R)$ denote a one-dimensional spatial domain, and $x_R\in\mathbb{R_+}$ is the length of the spatial domain in the $x$-direction.
We discretize the spatial domain $\Omega_x$ using $n_x$ grid points uniformly distributed with spacing $\Delta x=x_R/n_x$, where $n_x\in\mathbb{R}_+$ is the number of computational points in the $x$-direction.
Then, the nonlinear diffusion-reaction equation is discretized as
\begin{equation}
	\frac{\mathrm{d}\bm{\phi}(t)}{\mathrm{d}t}=(D\bm{\Delta}+Q)\bm{\phi}(t)+R\bm{\phi}(t)^2,	
	\label{eq:allen_cahn_discrete}
\end{equation}
where $\bm{\phi}(t)=\bm{\phi}=[\phi(t,x_0),\phi(t,x_1),\dots,\phi(t,x_{n_x-1})]^{\mathrm{T}}$ is the discretized field variable, $x_j$ for $j=0,1,\ldots,n_x-1$ indicates the spatial coordinates of the $j$-th node of the $x$-direction, and $\bm{\Delta}$ is the Laplace operator discretized by the second-order central difference method.
For notational convenience, we use $\phi_j$ given by $\phi_j\coloneqq\phi(t,x_j)$.
Given the Dirichlet boundary conditions $\phi_{-1}=\phi_{n_x}=0$, the Laplace operator $\bm{\Delta}$ discretized by the second-order central difference method is as follows:
\begin{equation}
	\bm{\Delta}=\frac{1}{(\Delta x)^2}\left[\begin{array}{cccc}
		-2 & 1 & & \\
		1 & -2 & 1 & \\
		& \ddots & \ddots & \ddots \\
		& & 1 & -2
	\end{array}\right].
\end{equation}
Eq.~\eqref{eq:allen_cahn_discrete} is also written as
\begin{equation}
    \frac{\mathrm{d}\bm{\phi}}{\mathrm{d}t}=\bm{F}_1\bm{\phi}+\bm{F}_2\bm{\phi}^{\otimes 2},
    \label{eq:allen_cahn_tensor}
\end{equation}
where $\bm{F}_1=D\bm{\Delta}+Q$ and $\bm{F}_2$ is a linear mapping of $\bm{\phi}^{\otimes 2}$ to $R\bm{\phi}_j^2$ for $j=0,1,\ldots,n_x-1$.

Subsequently, consider converting Eq.~\eqref{eq:allen_cahn_tensor} into a linear ODE using CL.
Let $\bm{\Phi}_k(t)= \bm{\phi}^{\otimes k}(t)$ and transform it into the following infinite-dimensional linear differential equation:
\begin{equation}
	\frac{\mathrm{d}\bm{\Phi}_k(t)}{\mathrm{d}t}=\bm{A}_{k,k}\bm{\Phi}_k(t)+\bm{A}_{k,k+1}\bm{\Phi}_{k+1}(t).
  \label{eq:allen_cahn_tensor_linear_finite}
\end{equation}
where $\bm{A}_{k,k}$ and $\bm{A}_{k,k-1}$ are respectively expressed as
\begin{align}
  \bm{A}_{k,k}&=\sum_{v=0}^{k-1}\bm{I}^{\otimes v}\otimes \bm{F}_{1} \otimes \bm{I}^{\otimes k-1-v},\\
  \bm{A}_{k,k+1}&=\sum_{v=0}^{k-1}\bm{I}^{\otimes v}\otimes \bm{F}_{2} \otimes \bm{I}^{\otimes k-1-v}.
\end{align}
Since Eq.~\eqref{eq:allen_cahn_tensor_linear_finite} is not solved in infinite dimensions, we truncate Eq.~\eqref{eq:allen_cahn_tensor_linear_finite} at order $K$ to obtain a finite-dimensional approximation.
The approximate finite-dimensional linear differential equation truncated Eq.~\eqref{eq:allen_cahn_tensor_linear_finite} at order $K$ is
\begin{equation}
	\frac{\mathrm{d}\bm{\Phi}(t)}{\mathrm{d}t}=\bm{A}{\bm{\Phi}(t)},
	\label{eq:Allen-Cahn_linearized}
\end{equation}
where $\bm{\Phi}$ and $\bm{A}$ are respectively defined as
\begin{equation}
	\bm{\Phi}(t)=[\bm{\Phi}_1(t),\bm{\Phi}_2(t),\dots,\bm{\Phi}_K(t)]^{\mathrm{T}},
\end{equation}
\begin{equation}
	\bm{A}=\left[\begin{array}{ccccc}
		{\bm A}_{1, 1} & {\bm A}_{1, 2} &  &  &\\
		& {\bm A}_{2, 2}& {\bm A}_{2, 3}&  &\\
		& & \ddots & \ddots &\\
		& & &{\bm A}_{K-1, K-1}& {\bm A}_{K-1, K}\\
		& & & &{\bm A}_{K, K}
	\end{array}\right].
\end{equation}
Matrix $\bm{A}$ is called the Carleman matrix \cite{Sanavio2024circuit} and vector $\bm{\Phi}$ is called the Carleman state vector.
The Carleman matrix is always a square matrix.
Next, Eq.~\eqref{eq:Allen-Cahn_linearized} is transformed into the Schr\"{o}dinger equation using WPT.
The Carleman matrix in Eq.~\eqref{eq:Allen-Cahn_linearized} is decomposed into its Hermitian part $\bm{H}_1$ and skew-Hermitian part $i\bm{H}_2$.
Then, the Carleman matrix that is not a skew-Hermitian matrix is
\begin{equation}
  \bm{A}=\bm{H}_1+i\bm{H}_2.
\end{equation}
Introduce a new auxiliary variable $p\geq0$ into the space variable $x$ of the system.
As described in Sec.~\ref{sec:WPT}, WPT is expressed as
\begin{equation}
  \bm{\psi}(t,p)=e^{-p}\bm{\Phi}(t).
\end{equation}
According to Eq.~\eqref{eq:linear_transformed_wpt}, this variable $\bm{\psi}(t,p)$ satisfies the following equation:
\begin{equation}
  \frac{\mathrm{d}\bm{\psi}(t,p)}{\mathrm{d}t}=\left(-\bm{H}_1\frac{\partial}{\partial p}+i\bm{H}_2\right)\bm{\psi}(t,p).
  \label{eq:cls_psi}
\end{equation} 
Let $\Omega_p\coloneqq(p_L, p_R),~p_L<p_R$ denote a one-dimensional domain, and $p_L$ and $p_R$ be the endpoints of the domain in the $p$-direction.
We discretize the spatial domain $\Omega_p$ using $n_p$ grid points uniformly distributed with spacing $\Delta p=(p_R-p_L)/n_p$, where $n_p\in\mathbb{R}_+$ is the number of computational points in the $p$-direction.
$p_j$ for $j=0,1,\ldots,n_p-1$ indicates the spatial coordinates of the $j$-th node of the $p$-direction.
We define the following vector:
\begin{equation}
  \bm{p}\coloneqq[e^{-p_0},e^{-p_1},\ldots,e^{-p_{n_p-1}}]^{\mathrm{T}}\in\mathbb{R}^{n_p}.
\end{equation}
Subsequently, we define $\bm{\psi}(t)$ as
\begin{equation}
	\bm{\psi}(t)\coloneqq\bm{p}\otimes\bm{\Phi}(t),
  \label{eq:wpt_psi}
\end{equation}
where $\bm{\psi}(t)=[\bm{\psi}(t,p_0),\bm{\psi}(t,p_1),\ldots,\bm{\psi}(t,p_{n_p-1})]^{\mathrm{T}}$.
For notational convenience, we use $\bm{\psi}_j(t)$ given by $\bm{\psi}_j(t)\coloneqq\bm{\psi}(t,p_j)$.
The variable $\bm{\psi}_j$ can be expressed as $\bm{\psi}_j(t)=e^{-p_j}\bm{\Phi}(t)$ for $j=0,1,\ldots,n_p-1$.
The first-order upwind difference method is used for the $p$-direction as the difference scheme in Eq.~\eqref{eq:cls_psi}.
The upwind difference method is a discretization technique in which the spatial differential term is approximated by the difference between a reference point and an upstream point.
Since Eq.~\eqref{eq:cls_psi} advects in the $p$-negative direction, applying the upwind difference method yields
\begin{equation}
	\frac{\mathrm{d}\bm{\psi}_j(t)}{\mathrm{d} t}=-\bm{H}_1\frac{\bm{\psi}_{j+1}(t)-\bm{\psi}_{j}(t)}{\Delta p}+i\bm{H}_2\bm{\psi}_j(t).
	\label{eq:cls_advection_scheme}
\end{equation}
In this study, the time evolution of the solution of the nonlinear reaction--diffusion equation by CLS is obtained by time evolving Eq.~\eqref{eq:cls_advection_scheme}.

We consider another representation of Eq.~\eqref{eq:cls_advection_scheme}.
Applying a left tensor product with $\bm{p}$ to both sides of Eq.~\eqref{eq:Allen-Cahn_linearized} yields
\begin{equation}
  \frac{\mathrm{d}}{\mathrm{d}t}(\bm{p}\otimes\bm{\Phi}(t))=\bm{p}\otimes\bm{H}_1\bm{\Phi}(t)+\bm{p}\otimes i\bm{H}_2\bm{\Phi}(t).
  \label{eq:allen_cahn_tensor_linear_finite_p}
\end{equation}
Eq.~\eqref{eq:allen_cahn_tensor_linear_finite_p} can be transformed into
\begin{equation}
  \frac{\mathrm{d}}{\mathrm{d}t}(\bm{p}\otimes\bm{\Phi}(t))=-\bm{\nabla}_p\bm{p}\otimes\bm{H}_1\bm{\Phi}(t)+\bm{p}\otimes i\bm{H}_2\bm{\Phi}(t),
  \label{eq:allen_cahn_tensor_linear_finite_p_transformed}
\end{equation}
where $\bm{\nabla}_p$ is the gradient operator in the $p$-direction discretized by the first-order upwind difference method.
Given the periodic boundary condition, considering the advection from right to left in the $p$ domain, the upwind difference method is defined as
\begin{equation}
  \bm{\nabla}_p=\frac{1}{\Delta p}
  \begin{bmatrix}
    -1 & 1 & & & \\
    & -1 & 1 & & \\
    &  & \ddots & \ddots & \\
    &  & & -1 & 1\\
    1 &  &  &  & -1
  \end{bmatrix}.
\end{equation}
Eq.~\eqref{eq:allen_cahn_tensor_linear_finite_p_transformed} can be transformed into
\begin{equation}
  \frac{\mathrm{d}}{\mathrm{d}t}(\bm{p}\otimes\bm{\Phi}(t))=(-\bm{\nabla}_p\otimes\bm{H}_1+\bm{I}\otimes i\bm{H}_2)(\bm{p}\otimes\bm{\Phi}(t)).
  \label{eq:allen_cahn_tensor_linear_finite_p_transformed2}
\end{equation}
Substituting Eq.~\eqref{eq:wpt_psi} into Eq.~\eqref{eq:allen_cahn_tensor_linear_finite_p_transformed2} yields
\begin{equation}
  \frac{\mathrm{d}\bm{\psi}}{\mathrm{d}t}=\tilde{\bm{H}}\bm{\psi},
  \label{eq:cls_advection_tensor}
\end{equation}
where
\begin{equation}
  \tilde{\bm{H}}\coloneqq-\bm{\nabla}_p\otimes\bm{H}_1+\bm{I}\otimes i\bm{H}_2.
  \label{eq:cls_advection_tensor_H}
\end{equation}
Eqs.~\eqref{eq:cls_advection_tensor} and \eqref{eq:cls_advection_tensor_H} are other representations of Eq.~\eqref{eq:cls_advection_scheme}.
When we consider the implementation of CLS on quantum circuits, the representations of Eqs.~\eqref{eq:cls_advection_tensor} and \eqref{eq:cls_advection_tensor_H} are more suitable than that of Eq.~\eqref{eq:cls_advection_scheme}.
We extend Eqs.~\eqref{eq:cls_advection_scheme} and \eqref{eq:cls_advection_tensor} to $p<0$ ($p_L<0, p_R>0$) with the following initial data:
\begin{equation}
  \bm{\psi}(0)=\bm{P}\otimes\bm{\Phi}(0),
\end{equation}
where
\begin{equation}
  \bm{P}=[e^{-|p_0|},e^{-|p_1|},\ldots,e^{-|p_{n_p-1}|}]^{\mathrm{T}}\in\mathbb{R}^{n_p}.
\end{equation}

\subsection{Classical numerical methods for CLS}
The discretization in the $x$- and $p$-directions discussed in Section \ref{sec:cls}, which was originally in the context of classical computation, can also be applied when implementing the method on a quantum computer.
Since Hamiltonian simulation operates analogously rather than digitally, it does not update the state step-by-step in time during simulations; instead, it can directly generate the dynamics corresponding to the desired time evolution in an analog manner.
In other words, when considering the implementation of CLS on a quantum computer, the time step size in classical simulation can effectively be set to zero.
Note that, while this eliminates the time step error inherent in classical simulation, it does not render the quantum simulation error-free.
Other types of error may arise depending on the quantum algorithm employed, such as time discretization errors in Suzuki–Trotter decompositions or approximation errors in the construction of time-evolution operators via quantum singular value transformation \cite{Martyn2021Grandunification}.

However, in this study, the primary objective is to investigate the utility and characteristics of CLS through classical simulations.
Therefore, for classical simulations, we consider discretization in the time direction.
Specifically, we examine the time evolution from the initial time $t = 0$ up to the integration time $t = T$.
Letting $n_t$ denote the number of time steps, we can express the time step size $\Delta t$ as $\Delta t = T / n_t$.
By applying the first-order forward difference scheme to the time derivative term in Eq.~\eqref{eq:cls_advection_scheme}, we obtain the following expression:
\begin{align}
  \frac{\bm{\psi}_j{(n+1)}-\bm{\psi}_j{(n)}}{\Delta t}=-\bm{H}_1\frac{\bm{\psi}_{j+1}{(n)}-\bm{\psi}_{j}{(n)}}{\Delta p}+i\bm{H}_2\bm{\psi}_j{(n)}.
  \label{eq:cls_advection_scheme_discrete}
\end{align}
Here, $n=0,1,\ldots,n_t-1$ is the number of time steps and $\bm{\psi}_j(n)$ represents $\bm{\psi}_j(t)$ at $t=n\Delta t$.
Rearranging with respect to $\bm{\psi}_j(n)$, we obtain the following equation:
\begin{align}
  \bm{\psi}_j{(n+1)}=&-\bm{H}_1\frac{\Delta t}{\Delta p}\bm{\psi}_{j+1}{(n)} \nonumber \\
  &+\left(1+\bm{H}_1\frac{\Delta t}{\Delta p}+i\bm{H}_2\Delta t\right)\bm{\psi}_j{(n)}.
  \label{eq:cls_advection_scheme_discrete2}
\end{align}
By introducing $\bm{B}_1$ and $\bm{B}_2$, we can rewrite the equation as
\begin{equation}
  \bm{\psi}_j{(n+1)}=\bm{B}_1\bm{\psi}_{j+1}{(n)}+\bm{B}_2\bm{\psi}_j{(n)},
  \label{eq:cls_advection_scheme_discrete4}
\end{equation}
where
\begin{equation}
  \bm{B}_1\coloneqq -\bm{H}_1\frac{\Delta t}{\Delta p},\quad \bm{B}_2\coloneqq 1+\bm{H}_1\frac{\Delta t}{\Delta p}+i\bm{H}_2\Delta t.
  \label{eq:cls_advection_scheme_discrete3}
\end{equation}
Given the boundary condition $\bm{\psi}_0=\bm{\psi}_{n_p}$, the final iterative system is as follows:
\begin{equation}
  \bm{\psi}(n+1)=\bm{B}\bm{\psi}(n),\quad n=0,1,\ldots,n_t-1,
  \label{eq:cls_advection_scheme_discrete_final}
\end{equation}
where
\begin{equation}
  \bm{B}=
  \begin{bmatrix}
    \bm{B}_2 & \bm{B}_1 & & & \\
    & \bm{B}_2 & \bm{B}_1 & & \\
    &  & \ddots & \ddots & \\
    &  & & \bm{B}_2 & \bm{B}_1\\
    \bm{B}_1&  &  &  & \bm{B}_2
  \end{bmatrix}.
\end{equation}
By iteratively computing Eq.~\eqref{eq:cls_advection_scheme_discrete_final}, we can obtain the solution vector of $\bm{\psi}$ at the desired time.

\begin{table*}[t] 
    \centering
    \caption{Computational conditions for FDM, CL, and CLS.}
    \label{tab:condition}
    \begin{tabular}{cc} 
      \hline\hline
      Computation domain of $x$& $x\in(0\mathrm{\,m},1\mathrm{\,m})$\\
      Computation domain of $p$& $p\in[p_L,p_R]=[-20\mathrm{\,m},20\mathrm{\,m}]$\\
      Number of calculation points for $x$ & $n_x=36$\\
      Number of calculation points for $p$& $n_p=256$\\
      Discretization method for $x$ & Second-order central difference method\\
      Discretization method for $p$ & First-order upwind difference method\\
      Time step size $\varDelta t$ & $\varDelta t=1.0\times 10^{-6}\mathrm{\,s}$\\
      Number of time steps $n_t$& $n_t=0.4 \times 10^6$\\
      Initial distribution $\phi(0,x)$ & $\phi(0,x)=0.5-0.5\cos(2\pi x)$\\
      Boundary condition for $x$ & Dirichlet condition ($\phi_{-1}=\phi_{n_x}=0$)\\
      Boundary condition for $p$ & Periodic condition ($\bm{\psi}_0=\bm{\psi}_{n_p}$)\\
      CL truncation order $K$ & 3\\
      Variables in equation $P,~Q,~R$& $P=1,~Q=1,~R=-1$\\
      \hline\hline
    \end{tabular}
  \end{table*}

\section{Results and discussion}\label{sec2}
The nonlinear reaction--diffusion equation shown in Eq.~\eqref{eq:reaction_diffusion} is analyzed by the proposed method CLS shown in Fig.~\ref{fig:carleman_schrodingerization}.
In this study, to investigate the usefulness of CLS, a discretized and time-evolved version of the finite differential method (FDM) using the central difference method and a time-evolved version of the linearized equation using CL were prepared and compared. 
The validity of the proposed method was evaluated by determining the accuracy of calculation by CLS.
The computational conditions in this study are shown in Table~\ref{tab:condition}.

\begin{figure*}[t]
    \includegraphics[width=\linewidth]{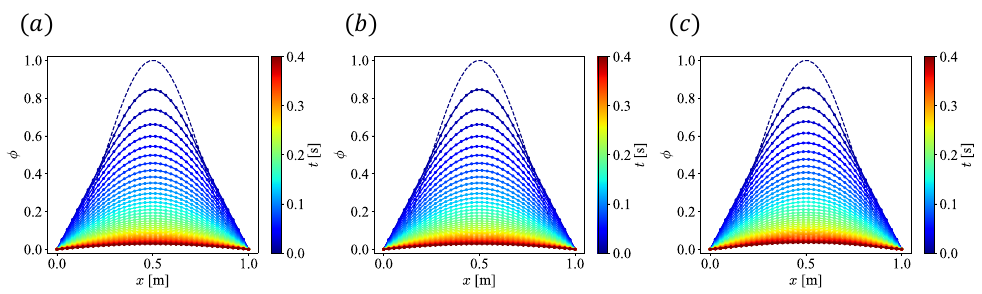}
    \caption{Time evolution of the solution of the nonlinear reaction--diffusion equation: (a) solution of FDM, (b) solution of CL, and (c) solution of CLS. Dashed lines represent the initial condition.}
    \label{fig:logistic_cos}
  \end{figure*}

\begin{figure*}[t]
    \centering
    \includegraphics[width=\linewidth]{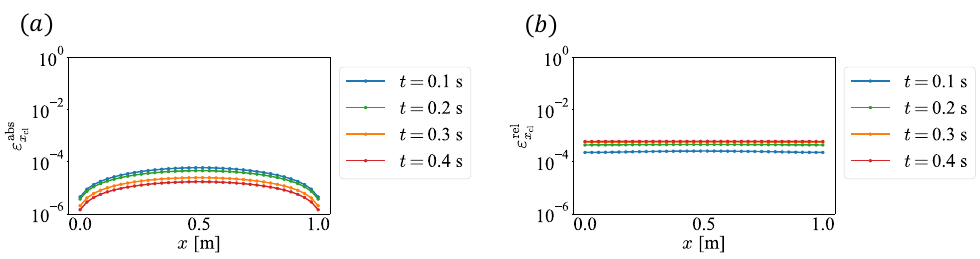}
    \caption{Error between CL and FDM based on the dynamics of the nonlinear reaction--diffusion equation: (a) absolute error $\varepsilon_{x_{\text{cl}}}^{\text{abs}}$ and (b) relative error $\varepsilon_{x_{\text{cl}}}^{\text{rel}}$.}
    \label{fig:logistic_cos_cl_fdm_error}
\end{figure*}

\begin{figure*}[t]
    \centering
    \includegraphics[width=\linewidth]{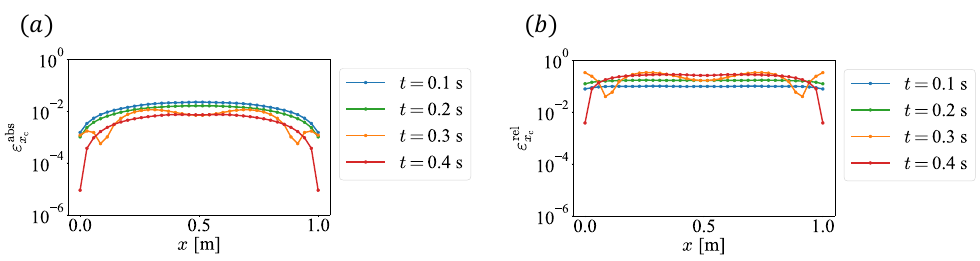}
    \caption{Error between CL and CLS: (a) absolute error $\varepsilon_{x_{\text{c}}}^{\text{abs}}$ and (b) relative error $\varepsilon_{x_{\text{c}}}^{\text{rel}}$.}
    \label{fig:logistic_cos_cl_cls_error}
\end{figure*}

\begin{figure*}[t]
    \centering
    \includegraphics[width=\linewidth]{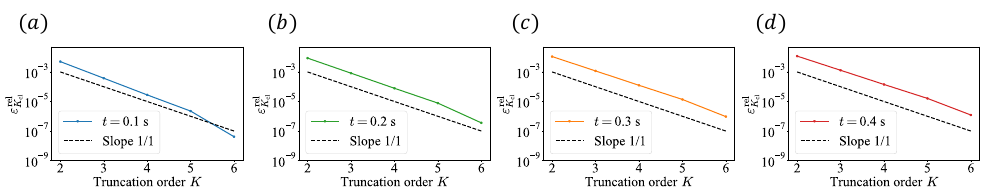}
    \caption{Relationship between the truncation order of CL and the relative error for the nonlinear reaction--diffusion equation solved using CL: (a) relative error at $t=0.1\mathrm{\,s}$, (b) relative error at $t=0.2\mathrm{\,s}$, (c) relative error at $t=0.3\mathrm{\,s}$, and (d) relative error at $t=0.4\mathrm{\,s}$.}
    \label{fig:truncate_cl_rel}
\end{figure*}

\begin{figure}[t]
    \centering
    \includegraphics[width=\linewidth]{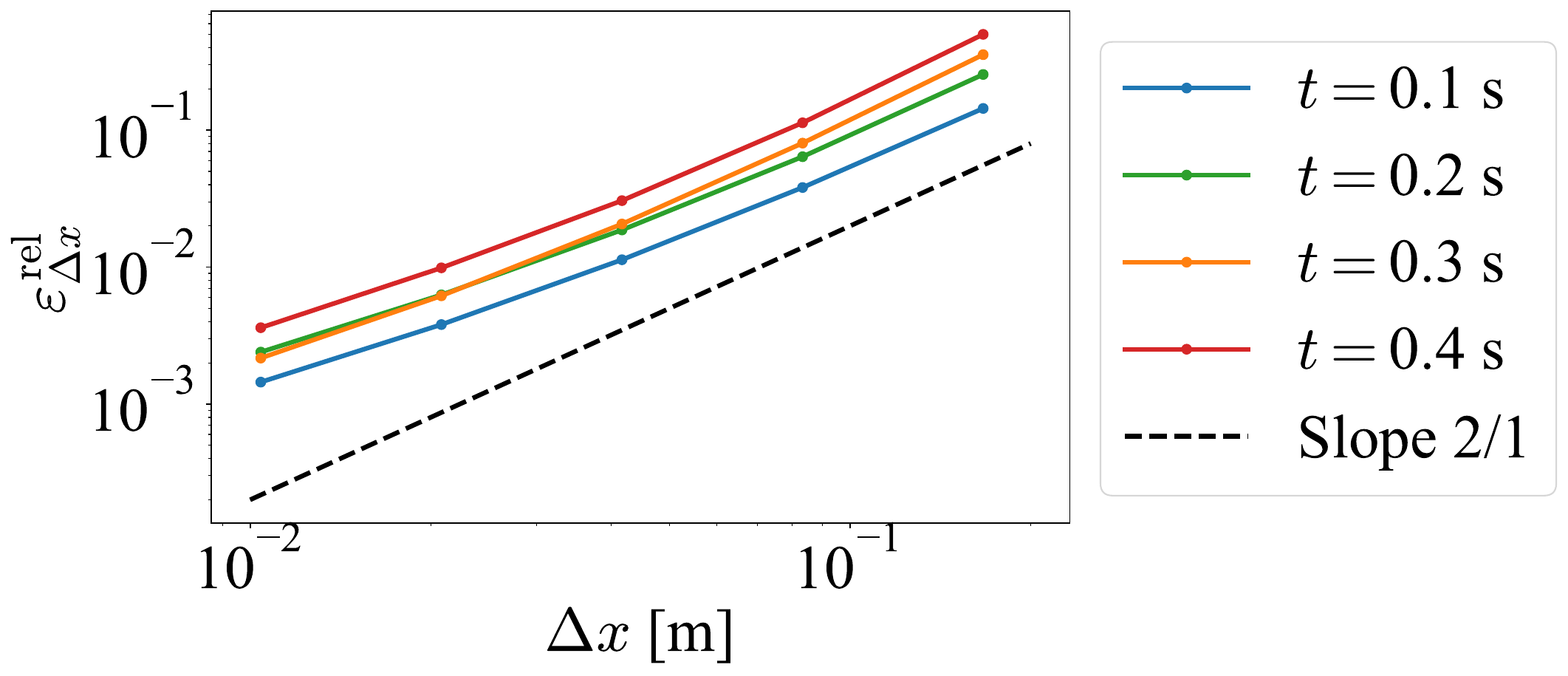}
    \caption{Computational accuracy of CLS with respect to $\varDelta x$.}
    \label{fig:deltax_cls}
\end{figure}

\begin{figure}[t]
    \centering
    \includegraphics[width=\linewidth]{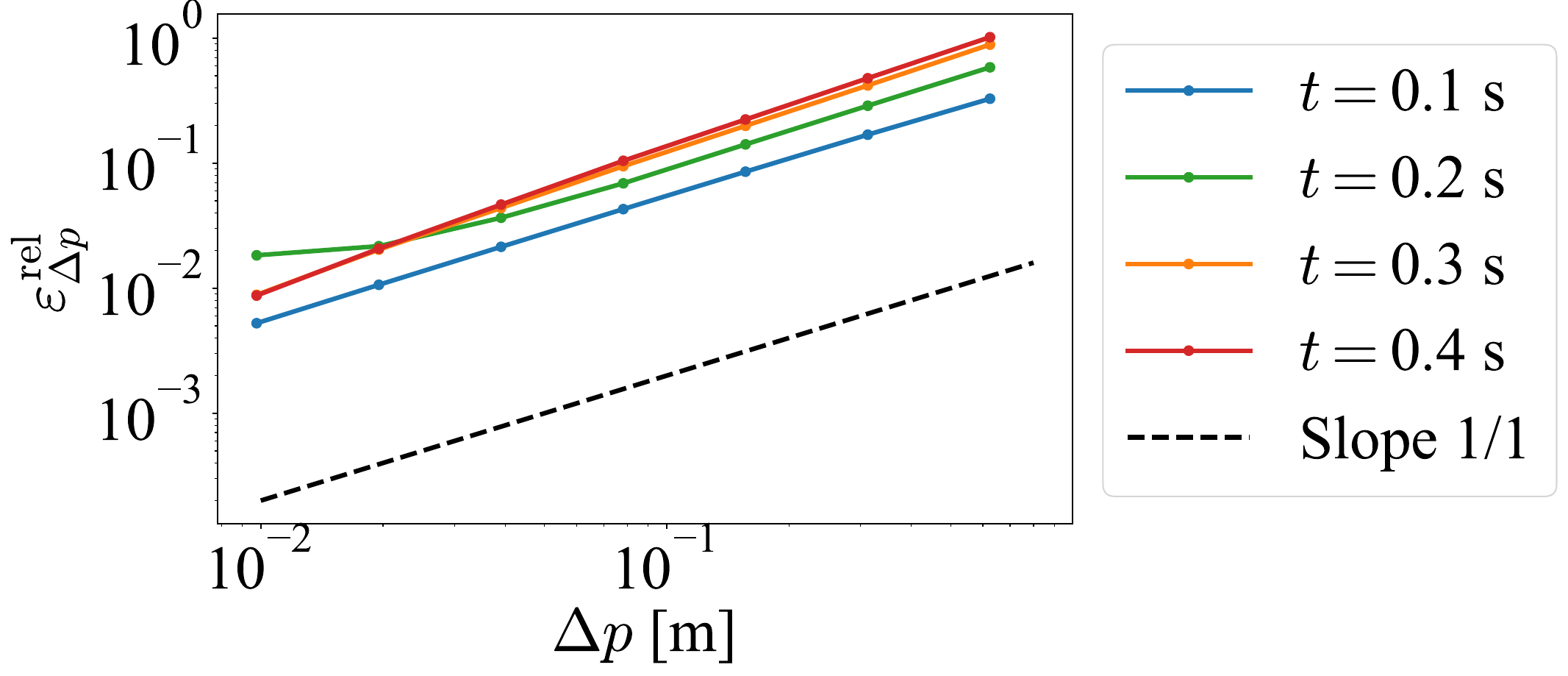}
    \caption{Computational accuracy of CLS with respect to $\Delta p$.}
    \label{fig:perror}
\end{figure}

\begin{figure}[t]
\centering
\includegraphics[width=\linewidth]{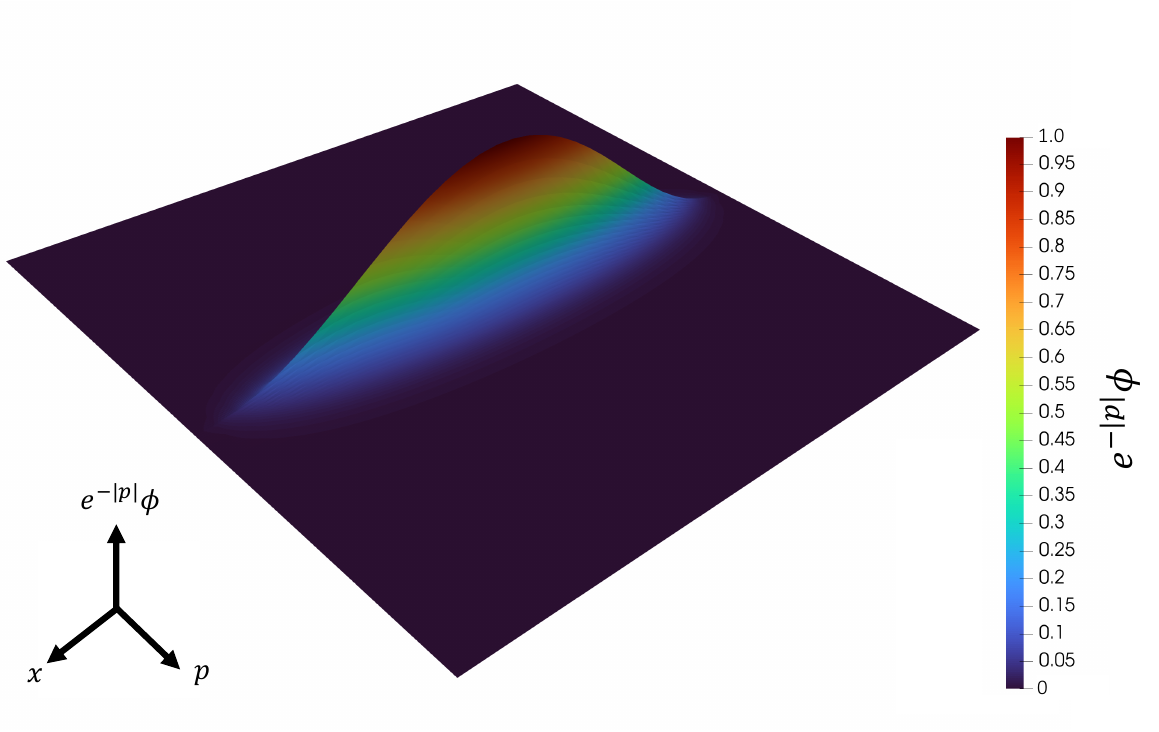}
\caption{3D visualization of initial distribution $\bm{\psi}(0)$.}
\label{fig:3d_wpt}
\end{figure}

\begin{figure*}[t]
  \centering
  \includegraphics[width=\linewidth]{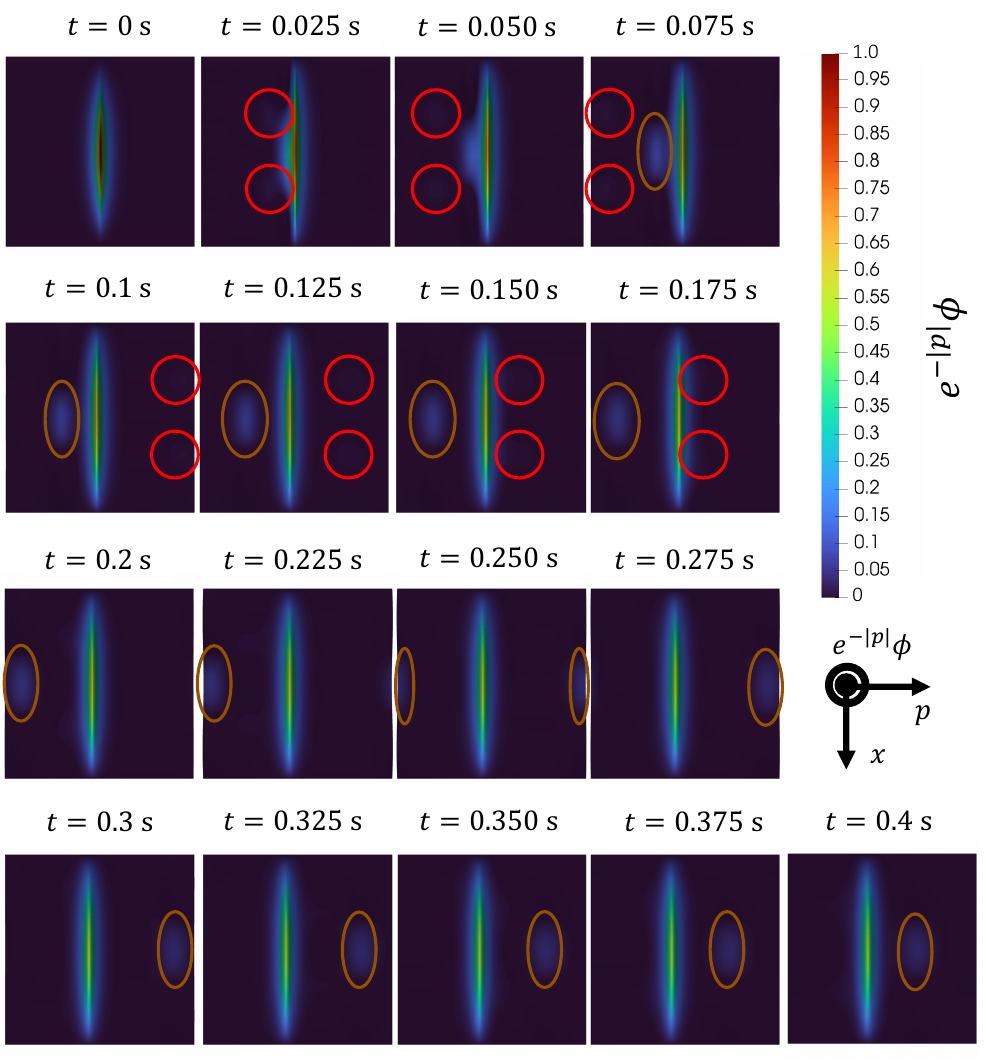}
  \caption{Time evolutions of the solutions by CLS and conventional method.}
  \label{fig:advection_evolution_time}
\end{figure*}

\subsection{Time evolution of the solution by CLS}
The time evolution of the solutions of Eq.~\eqref{eq:reaction_diffusion_transformed} by FDM, CL and CLS are shown in Figs.~\ref{fig:logistic_cos}(a)--(b) and 3(c), respectively.
It can be seen that the time evolution of the solution by the proposed method CLS qualitatively coincides with those of the solutions by FDM and CL in Figs.~\ref{fig:logistic_cos}(a) and 3(b).

\subsection{Accuracy of CLS calculations}
In this section, we discuss the accuracy of CLS calculations.
The error of CL with FDM as the true value is shown in Fig.~\ref{fig:logistic_cos_cl_fdm_error}, and that of CLS with CL as the true value is shown in Fig.~\ref{fig:logistic_cos_cl_cls_error}.
From Figs.~\ref{fig:logistic_cos_cl_fdm_error} and \ref{fig:logistic_cos_cl_cls_error}, we see that the error due to CLS is larger than that due to CL.
Therefore, it can be assumed that the error due to CLS is dominated by that introduced by the WPT process.

Next, we investigated the accuracy of the calculation for the truncated order $K$ of CL.
The relative error of CL when FDM is taken as the true value for the truncated order $K$ of CL is shown in Fig.~\ref{fig:truncate_cl_rel}.
From Fig.~\ref{fig:truncate_cl_rel}, we see that the relative error is parallel to the line with slope 1/1 on both logarithmic plots.
Therefore, we can consider that CL is first-order-accurate for the truncated order $K$.

Next, the accuracy of the calculation is investigated for the $x$-direction.
The relative error of CLS when FDM is taken as the true value for the spatial step size in the $x$-direction, $\Delta x$, is shown in Fig.~\ref{fig:deltax_cls}.
From Fig.~\ref{fig:deltax_cls}, we see that the relative error is parallel to the straight line with slope of 2/1 on both logarithmic plots.
CLS shows second-order accuracy for the $x$ spatial step size $\Delta x$.
This result implies that it is based on the discretization using a second-order central difference method for the $x$-direction.

Next, we examine the accuracy of the calculation for the $p$-direction.
The relative error of CLS when CL is the true value for CLS with respect to the spatial step size in the $p$-direction $\Delta p$ is shown in Fig.~\ref{fig:perror}.
From Fig.~\ref{fig:perror}, we see that the relative error is parallel to the line with slope 1 on both logarithmic plots.
Therefore, we can consider that CLS is first-order-accurate for the $p$ spatial step size $\Delta p$.
This result implies that it is based on the discretization using the first-order upwind difference method for the $p$-direction.

We consider the error caused by advection in WPT.
In WPT, the solution advects in the direction of $p<0$.
Fig.~\ref{fig:3d_wpt} shows a 3D plot of the initial distribution in WPT.
Fig.~\ref{fig:advection_evolution_time} shows the time evolution of the solution obtained by WPT on the $xp$ plane.
The red circle surrounds the wave that first affects the calculation accuracy.
The brown circle surrounds the wave that next affects the calculation accuracy.
Note that periodic boundary conditions are imposed in the $p$-direction.
The red and brown circles have different propagation speeds.
Although the computational domain is extended to the $p < 0$ region under the assumption that it does not affect the solution in the $p \geq 0$ region, it is considered that this affects the computational accuracy because it affect the $p \geq 0$ region.
For long-term simulations, the errors caused by this advection in WPT can be a problem.

\section{Conclusion}\label{sec:conclusion}
In this study, the proposed CLS method was applied to the nonlinear reaction--diffusion equation, and the time evolution of the solution by CLS and its computational accuracy were investigated.
The time evolution of the solution by CLS was almost the same as that by the conventional method.
The computational accuracy of CLS was found to be first-order accuracy for the truncated order of CL, second-order accuracy for the spatial variable $x$-direction, and first-order accuracy for the auxiliary variable $p$-direction.
The computational accuracy in the $x$- and $p$-directions was considered to be the result of discretization by the second-order accuracy central difference and first-order accuracy upwind difference methods, respectively.
This indicated that the computations performed using CLS are consistent with the theoretical predictions.
The proposed CLS method extended the framework of time evolution simulation in quantum computing and newly shows that Hamiltonian simulation can be applied to nonlinear PDEs.

\begin{acknowledgments}
This work was supported by JST FOREST Program, Japan (Grant Number JPMJFR212K).
\end{acknowledgments}
\bibliographystyle{unsrt}
\bibliography{reference}

\begin{thebibliography}{10}

\bibitem{Sato2024scalableqc}
Yuki Sato, Ruho Kondo, Ikko Hamamura, Tamiya Onodera, and Naoki Yamamoto.
\newblock Hamiltonian simulation for hyperbolic partial differential
  equations^^c2^^a0by scalable quantum circuits.
\newblock {\em Physical Review Research}, 6:033246, 9 2024.

\bibitem{Sato2024LCHS}
Yuki Sato, Hiroyuki Tezuka, Ruho Kondo, and Naoki Yamamoto.
\newblock Quantum algorithm for partial differential equations of
  non-conservative systems with spatially varying parameters.
\newblock {\em arXiv}, 7 2024.

\bibitem{Gibney2019quantumsupremacy}
Elizabeth Gibney.
\newblock Hello quantum world! google publishes landmark quantum supremacy
  claim.
\newblock {\em Nature}, 574:461--462, 10 2019.

\bibitem{Sukulthanasorn2025}
Naruethep Sukulthanasorn, Junsen Xiao, Koya Wagatsuma, Reika Nomura, Shuji
  Moriguchi, and Kenjiro Terada.
\newblock A novel design update framework for topology optimization with
  quantum annealing: Application to truss and continuum structures.
\newblock {\em Computer Methods in Applied Mechanics and Engineering},
  437:117746, 3 2025.

\bibitem{LiuOrtiz2023}
Burigede Liu, Michael Ortiz, and Fehmi Cirak.
\newblock Towards quantum computational mechanics.
\newblock {\em Computer Methods in Applied Mechanics and Engineering},
  432:117403, 12 2023.

\bibitem{Sarma2024}
Abhijat Sarma, Thomas~W. Watts, Mudassir Moosa, Yilian Liu, and Peter~L.
  McMahon.
\newblock Quantum variational solving of nonlinear and multidimensional partial
  differential equations.
\newblock {\em Physical Review A}, 109:062616, 6 2024.

\bibitem{Costa2022}
Pedro~C.S. Costa, Dong An, Yuval~R. Sanders, Yuan Su, Ryan Babbush, and
  Dominic~W. Berry.
\newblock Optimal scaling quantum linear-systems solver via discrete adiabatic
  theorem.
\newblock {\em PRX Quantum}, 3:040303, 10 2022.

\bibitem{Motlagh2023}
Danial Motlagh and Nathan Wiebe.
\newblock Generalized quantum signal processing.
\newblock {\em PRX Quantum}, 5, 4 2024.

\bibitem{Childs2020}
Andrew~M. Childs and Jin-Peng Liu.
\newblock Quantum spectral methods for differential equations.
\newblock {\em Communications in Mathematical Physics}, 375:1427--1457, 4 2020.

\bibitem{Berry2017}
Dominic~W. Berry, Andrew~M. Childs, Aaron Ostrander, and Guoming Wang.
\newblock Quantum algorithm for linear differential equations with
  exponentially improved dependence on precision.
\newblock {\em Communications in Mathematical Physics}, 356:1057--1081, 1 2017.

\bibitem{Berry2014}
Dominic~W Berry.
\newblock High-order quantum algorithm for solving linear differential
  equations.
\newblock {\em Journal of Physics A: Mathematical and Theoretical}, 47:105301,
  3 2014.

\bibitem{Childs2017}
Andrew~M. Childs, Robin Kothari, and Rolando~D. Somma.
\newblock Quantum algorithm for systems of linear equations with exponentially
  improved dependence on precision.
\newblock {\em SIAM Journal on Computing}, 46:1920--1950, 1 2017.

\bibitem{Harrow2009HHL}
Aram~W. Harrow, Avinatan Hassidim, and Seth Lloyd.
\newblock Quantum algorithm for linear systems of equations.
\newblock {\em Physical Review Letters}, 103:150502, 10 2009.

\bibitem{Dervovic2018}
Danial Dervovic, Mark Herbster, Peter Mountney, Simone Severini, Na^^c3^^afri
  Usher, and Leonard Wossnig.
\newblock Quantum linear systems algorithms: a primer.
\newblock {\em arXiv preprint}, 2 2018.

\bibitem{Childs2017HHLOPTIMAL}
Andrew~M. Childs, Robin Kothari, and Rolando~D. Somma.
\newblock Quantum algorithm for systems of linear equations with exponentially
  improved dependence on precision.
\newblock {\em SIAM Journal on Computing}, 46:1920--1950, 2017.

\bibitem{Ye2023}
Linlin Ye, Zhaoqi Wu, and Shao~Ming Fei.
\newblock Coherence dynamics in quantum algorithm for linear systems of
  equations.
\newblock {\em Physica Scripta}, 98, 12 2023.

\bibitem{Pilaszewicz2025}
Cezary Pilaszewicz and Marian Margraf.
\newblock Black-box security of stream ciphers under the quantum algorithm for
  linear systems of equations.
\newblock {\em Discover Computing}, 28:31, 4 2025.

\bibitem{Low2017}
Guang~Hao Low and Isaac~L. Chuang.
\newblock Optimal hamiltonian simulation by quantum signal processing.
\newblock {\em Physical Review Letters}, 118:010501, 1 2017.

\bibitem{Berry2015}
Dominic~W. Berry, Andrew~M. Childs, Richard Cleve, Robin Kothari, and
  Rolando~D. Somma.
\newblock Simulating hamiltonian dynamics with a truncated taylor series.
\newblock {\em Physical Review Letters}, 114:090502, 3 2015.

\bibitem{Childs2018}
Andrew~M. Childs, Dmitri Maslov, Yunseong Nam, Neil~J. Ross, and Yuan Su.
\newblock Toward the first quantum simulation with quantum speedup.
\newblock {\em Proceedings of the National Academy of Sciences},
  115:9456--9461, 11 2018.

\bibitem{Linlinlecture2022}
Lin Lin.
\newblock Lecture notes on quantum algorithms for scientific computation.
\newblock 1 2022.

\bibitem{Babbush2023quantumspeedup}
Ryan Babbush, Dominic~W. Berry, Robin Kothari, Rolando~D. Somma, and Nathan
  Wiebe.
\newblock Exponential quantum speedup in simulating coupled classical
  oscillators.
\newblock {\em Physical Review X}, 13:041041, 12 2023.

\bibitem{Joseph2020}
Ilon Joseph.
\newblock {Koopman-von Neumann approach to quantum simulation of nonlinear
  classical dynamics}.
\newblock {\em Physical Review Research}, 2(4), 2020.

\bibitem{Koopman1931}
B.~O. Koopman.
\newblock Hamiltonian systems and transformation in hilbert space.
\newblock {\em Proceedings of the National Academy of Sciences}, 17:315--318, 5
  1931.

\bibitem{Liu2021}
Jin-Peng Liu, Herman ^^c3^^98ie Kolden, Hari~K. Krovi, Nuno~F. Loureiro,
  Konstantina Trivisa, and Andrew~M. Childs.
\newblock Efficient quantum algorithm for dissipative nonlinear differential
  equations.
\newblock {\em Proceedings of the National Academy of Sciences}, 118, 8 2021.

\bibitem{Liu2023carlemanreactiondiffusion}
Jin-Peng Liu, Dong An, Di~Fang, Jiasu Wang, Guang~Hao Low, and Stephen Jordan.
\newblock Efficient quantum algorithm for nonlinear
  reaction^^e2^^80^^93diffusion equations and energy estimation.
\newblock {\em Communications in Mathematical Physics}, 404:963--1020, 12 2023.

\bibitem{Amini2025}
Arash Amini, Cong Zheng, Qiyu Sun, and Nader Motee.
\newblock Carleman linearization of nonlinear systems and its finite-section
  approximations.
\newblock {\em Discrete and Continuous Dynamical Systems - B}, 30:577--603,
  2025.

\bibitem{Akiba2023}
Takaki Akiba, Youhi Morii, and Kaoru Maruta.
\newblock Carleman linearization approach for chemical kinetics integration
  toward quantum computation.
\newblock {\em Scientific Reports}, 13:3935, 3 2023.

\bibitem{Forets2021}
Marcelo Forets and Christian Schilling.
\newblock {\em Reachability of Weakly Nonlinear Systems Using Carleman
  Linearization}, pages 85--99.
\newblock Springer International Publishing, 8 2021.

\bibitem{Forets2017}
Marcelo Forets and Amaury Pouly.
\newblock Explicit error bounds for carleman linearization.
\newblock {\em arXiv preprint}, 11 2017.

\bibitem{Sanavio2024}
Claudio Sanavio, Enea Mauri, and Sauro Succi.
\newblock Carleman-grad approach to the quantum simulation of fluids.
\newblock {\em arXiv preprint}, 6 2024.

\bibitem{Shi2024}
Dongwei Shi and Xiu Yang.
\newblock Koopman spectral linearization vs. carleman linearization: A
  computational comparison study.
\newblock {\em Mathematics}, 12, 7 2024.

\bibitem{Endo2024}
Katsuhiro Endo and Kazuaki~Z. Takahashi.
\newblock Divergence-free algorithms for solving nonlinear differential
  equations on quantum computers.
\newblock {\em arXiv preprint}, 11 2024.

\bibitem{Costa2019}
Pedro C.~S. Costa, Stephen Jordan, and Aaron Ostrander.
\newblock Quantum algorithm for simulating the wave equation.
\newblock {\em Physical Review A}, 99:012323, 1 2019.

\bibitem{Gonzalez-Conde2023}
Javier Gonzalez-Conde, {\'{A}}ngel Rodr{\'{i}}guez-Rozas, Enrique Solano, and
  Mikel Sanz.
\newblock {Efficient Hamiltonian simulation for solving option price dynamics}.
\newblock {\em Physical Review Research}, 5(4):043220, 12 2023.

\bibitem{An2023}
Dong An, Jin~Peng Liu, and Lin Lin.
\newblock {Linear Combination of Hamiltonian Simulation for Nonunitary Dynamics
  with Optimal State Preparation Cost}.
\newblock {\em Physical review letters}, 131(15):150603, 2023.

\bibitem{Childs2012}
Andrew~M. Childs and Nathan Wiebe.
\newblock Hamiltonian simulation using linear combinations of unitary
  operations.
\newblock {\em arXiv preprint}, 2 2012.

\bibitem{Childs2021}
Andrew~M. Childs, Yuan Su, Minh~C. Tran, Nathan Wiebe, and Shuchen Zhu.
\newblock Theory of trotter error with commutator scaling.
\newblock {\em Physical Review X}, 11, 2 2021.

\bibitem{Meister2022}
Richard Meister, Simon~C. Benjamin, and Earl~T. Campbell.
\newblock Tailoring term truncations for electronic structure calculations
  using a linear combination of unitaries.
\newblock {\em Quantum}, 6:637, 2 2022.

\bibitem{Jin2022}
Shi Jin, Nana Liu, and Yue Yu.
\newblock Quantum simulation of partial differential equations via
  schr^^c3^^b6dingerization.
\newblock {\em Physical Review Letters}, 133:230602, 12 2024.

\bibitem{Jin2023schodingerizaiton-first}
Shi Jin, Nana Liu, and Yue Yu.
\newblock Quantum simulation of partial differential equations: Applications
  and detailed analysis.
\newblock {\em Physical Review A}, 108:032603, 9 2023.

\bibitem{Jin2024schrodngerization-conditions}
Shi Jin, Xiantao Li, Nana Liu, and Yue Yu.
\newblock Quantum simulation for partial differential equations with physical
  boundary or interface conditions.
\newblock {\em Journal of Computational Physics}, 498:112707, 2 2024.

\bibitem{DeMasi1986}
A.~De Masi, P.~A. Ferrari, and J.~L. Lebowitz.
\newblock Reaction-diffusion equations for interacting particle systems.
\newblock {\em Journal of Statistical Physics}, 44:589--644, 8 1986.

\bibitem{Sanavio2024circuit}
Claudio Sanavio, Enea Mauri, and Sauro Succi.
\newblock Explicit quantum circuit for simulating the
  advection-diffusion-reaction dynamics.
\newblock {\em IEEE Transactions on Quantum Engineering}, 6:1--12, 2025.

\bibitem{Martyn2021Grandunification}
John~M. Martyn, Zane~M. Rossi, Andrew~K. Tan, and Isaac~L. Chuang.
\newblock Grand unification of quantum algorithms.
\newblock {\em PRX Quantum}, 2:040203, 12 2021.

\end{thebibliography}

\appendix
\section{Skew-Hermitianity of $\bm{A}'$} \label{sec:hermitian}
The operator $\bm{A}'$ obtained from WPT is
\begin{equation}
  \bm{A}'=-\bm{H}_1\frac{\partial}{\partial p}+i\bm{H}_2,
\end{equation}
where $\bm{H}_1$ is the Hermitian operator and $\bm{H}_2$ is the skew-Hermitian operator.
We define $\mathbb{R}_p=\{p\in \mathbb{R}\}$ and consider the following Hilbert space $\mathcal{H}$:
\begin{align}
  \mathcal{H}&=L^2(\mathbb{R}_p)\nonumber\\
  &=\left\{f:\mathbb{R}\rightarrow\mathbb{C}~\left|~\int_{-\infty}^{\infty}|f(p)|^2dp<\infty\right.\right\}.
\end{align}
At this time, the inner product is defined as
\begin{equation}
  \braket{\varphi,\psi}=\int_{-\infty}^{\infty}\varphi^*(p)\psi(p)dp,\quad \varphi,\psi\in\mathcal{H},
\end{equation}
where $\varphi^*(p)$ is the complex conjugate of $\varphi(p)$.
When an operator $\bm{G}$ satisfies the following equation, $\bm{G}$ is called a skew-Hermitian operator:
\begin{equation}
  \braket{\varphi,\bm{G}\psi}=-\braket{\bm{G}\varphi,\psi},\quad \varphi,\psi\in\mathcal{H}.
\end{equation}
Consider the following inner product to confirm the skew-Hermitian property of the operator $\partial/\partial p$:
\begin{align}
  \Braket{\varphi,\frac{\partial\psi}{\partial p}}&=\int_{-\infty}^{\infty}\varphi^*(p)\frac{\partial\psi(p)}{\partial p}dp,\nonumber\\
  &=\left[\varphi^*(p)\psi(p)\right]_{-\infty}^{\infty}-\int_{-\infty}^{\infty}\frac{\partial\varphi^*(p)}{\partial p}\psi(p)dp,\nonumber\\
  &=-\int_{-\infty}^{\infty}\left(\frac{\partial\varphi(p)}{\partial p}\right)^*\psi(p)dp,\nonumber\\
  &=-\Braket{\frac{\partial\varphi}{\partial p},\psi}.
\end{align}
Therefore, the operator $\partial/\partial p$ is a skew-Hermitian operator.
The complex transpose of operator $\bm{A}'$ is as follows:
\begin{align}
  \bm{A}'^\dagger&=\left(-\bm{H}_1\frac{\partial}{\partial p}+i\bm{H}_2\right)^\dagger,\nonumber\\
  &=\left(-\bm{H}_1\frac{\partial}{\partial p}\right)^\dagger+(i\bm{H}_2)^\dagger,\nonumber\\
  &=-\left(\frac{\partial}{\partial p}\right)^\dagger\bm{H}_1^\dagger-i\bm{H}_2,\nonumber\\
  &=\frac{\partial}{\partial p}\bm{H}_1-i\bm{H}_2,\nonumber\\
  &=-\left(\bm{H}_1\frac{\partial}{\partial p}+i\bm{H}_2\right),\nonumber\\
  &=-\bm{A}'.
\end{align}
Therefore, $\bm{A}'$ is a skew-Hermitian operator.

\end{document}